\documentclass[twocolumn,english,superscriptaddress,prl]{revtex4-1}
\usepackage[T1]{fontenc}
\usepackage[latin9]{inputenc}
\usepackage{float}
\usepackage{amsmath}
\usepackage{graphicx}
\usepackage[dvipsnames]{xcolor}

\makeatletter
\usepackage{babel}

\makeatother

\usepackage{babel}
\begin{document}
\title{Effect of directed aging on nonlinear elasticity and memory formation in a material
}
\author{Daniel Hexner}
\altaffiliation{Current address: Faculty of Mechanical Engineering, Technion, 320000 Haifa, Israel.} \email{danielhe@me.technion.ac.il}
\affiliation{Department of Physics and The James Franck and Enrico Fermi
Institutes, University of Chicago, Chicago IL, 60637.}
\affiliation{Department of Physics and Astronomy,University of Pennsylvania, Philadelphia PA, 19104.}
\thanks{DH and NP contributed equally to this work.}

\author{Nidhi Pashine}
\thanks{DH and NP contributed equally to this work.}
\affiliation{Department of Physics and The James Franck and Enrico Fermi
Institutes, University of Chicago, Chicago IL, 60637.}

\author{Andrea J. Liu}\affiliation{Department of Physics and Astronomy,University of Pennsylvania, Philadelphia PA, 19104.}
\author{Sidney R. Nagel}
\affiliation{Department of Physics and The James Franck and Enrico Fermi
Institutes, University of Chicago, Chicago IL, 60637.}

\begin{abstract}
Disordered solids often change their elastic response as they  slowly age. %We observe the effect of aging on the non-linear elastic response of materials. 
Using experiments and simulations, we study how aging %quasi-two-dimensional
disordered planar networks under an applied stress affects their nonlinear elastic response. We are able to modify dramatically the elastic properties of our systems in the non-linear regime. %The aged networks also show a memory of their aging history.
Using simulations, we study two models for the microscopic evolution of properties of such a material; the first considers changes in the material strength while the second considers distortions in the microscopic geometry. Both models capture different aspects of the experiments including the encoding of memories of the aging history of the system and the dramatic effects on the material's nonlinear elastic properties. Our results demonstrate how aging can be used to create complex elastic behavior in the nonlinear regime.
\end{abstract}

\maketitle
\section{Introduction}

The inevitable fate of a glass left on its own is to age and progressively lower its free energy in a rugged energy landscape~\cite{Struik_1977_Polymer,Hodge_1995_Science,mitchell2008aging,Hutchinson_1995_review}.  %as they evolve, towards equilibrium 
As it ages, particles rearrange or bonds break and form. As a result, the elastic properties evolve but, due to enormous
relaxation times, the system does not reach equilibrium at any accessible
time scale. Because preparation
into the initial metastable state can produce desired properties that are inaccessible in thermal equilibrium, aging is often considered to be detrimental
since it allows the system to evolve away from this state.

It was recently proposed, however, that stress-induced aging can be exploited to manipulate an out-of-equilibrium solid to achieve various desired elastic responses~\cite{pashine2019directed}.
Thus, imposing strain \emph{directs} the manner in which a solid
ages. This directed aging relies on the fact that straining a disordered system
gives rise to a spatially varying stress pattern that depends sensitively
on the applied deformation. 
If we consider a disordered network of nodes connected by bonds, an external force applied on such a network would result in a different stress at each bond. Each bond evolves (\textit{i.e.}, ages) at a different rate; in many cases, the bonds under the highest stress evolve the fastest. 

It had previously been shown that pruning the bonds according to the stress they feel under an applied deformation leads to novel and controllable elastic properties~\cite{goodrich2015principle}. 
%the elastic response of a disordered system depends sensitively on the precise microscopic details (stiffness and position) of the bonds. 
Replacing bond pruning by bond aging under an applied macroscopic strain gives rise to similar effects~\cite{pashine2019directed}.  This aging changes the corresponding stiffnesses or geometry of the individual bonds and thereby modifies the system's elastic moduli. In this way we have manipulated the elastic response in order to create auxetic (\textit{i.e.}, negative Poisson's ratio $\nu<0$) materials~\cite{lakes_r_Science_1987,pashine2019directed}. These auxetic materials are not ubiquitous in nature and are thought to have interesting
applications~\cite{RLakes_Science_1987,greaves2011poisson,ren2018auxetic,reid2019auxetic,Lakes_2017,YLiu_2010,yang2004review,alderson2007auxetic,evans2000auxetic,sanami2014auxetic,saxena2016three,liu2019realizing,rens2019rigidity}. In our simulations using oscillatory training, we have also been able to direct the aging process to achieve more local responses that mimic allosteric behavior in proteins~\cite{hexner2019periodic}.

In this paper, we focus on the \emph{nonlinear} elastic response of a system as it ages. %We find that aging under an applied stress dramatically affects its nonlinear response. %Here we show that a system can show a complex response in the nonlinear elastic regime simply by aging under a constant applied strain. This result demonstrates the power of the directed-aging protocol.
We find that directed aging can produce desired complex non-monotonic behaviors in the nonlinear elastic response of disordered networks.
%as well as in the linear response of a disordered network. We can attain complex non-monotonic behavior in the

We begin with data from aging experiments on physical networks. Even after the networks age under compression, they retain a distinct memory of their initial state. This memory can easily be discerned by measuring the response of these networks as a function of imposed strain. We show that elastic properties of these networks can be trained to have auxetic behavior at one value of strain and normal behavior at another.
Our results demonstrate the power of the directed aging protocol.

Earlier work has shown that at least two different factors (material weakening and geometry change) define the aging process in a disordered system~\cite{pashine2019directed} and that both are essential for fully understanding the process of directed aging. To isolate these effects, two different models were previously introduced~\cite{pashine2019directed}. One model weakens individual bonds in the material in response to stress and the other changes the network geometry of the bonds. We perform numerical simulations based on both models and show that while they produce intriguing differences, they retain a memory of how the material was trained. These results are in accord with our experimental findings.  %in the experiments are a robust feature of materials that are aged at high strains. 
These simulation results motivated further experiments in order to separate the material weakening effects from those of geometrical change; with these experiments we were able to accentuate the effects of material weakening while minimizing any change in geometry. %; the results are in excellent qualitative agreement with the simulations.

Finally, we characterize the inherent capacity of a network to be manipulated
as a function of strain, establishing that the system is more easily
trained under compression than expansion and that networks are more easily trained when they have a lower coordination number. 
Taken together, our results open the possibility of manipulating the elastic
properties of a material far beyond the linear regime.

\section{Experiments}
\label{experiments1}

We have recently shown that aging under various strains is an effective means of acquiring new elastic responses~\cite{pashine2019directed}. One such example is a negative Poisson's ratio, obtained by aging under compression (similar to the protocol of Ref. ~\cite{lakes_r_Science_1987}). In  previous work, we focused on how the type of deformation determines the resulting response.  In the present paper we focus on understanding the role of strain magnitude, aiming at understanding the nonlinear regime. Nonlinear effects induced by compression were previously studied in Ref.~\cite{choi1992non}.  Here we search for signatures of the aging history in the elastic response of the aged solid.

%When an external stress is applied on a disordered material, it results in a deformation of the system. If left to age under this applied stress, the disordered system would evolve in a way that relieves it of this stress. As a direct consequence of this, the energy required for the deformation goes down. This argument implies that aging a disordered system under uniform compression would lead to lowering the bulk modulus of the system. 

%We have recently shown that not only is this true, but it is also an effective way to make materials with a negative Poisson's ratio \cite{pashine2019directed} \color{red}using a similar protocol to that in Ref. ~\cite{lakes_r_Science_1987}}. In the previous work, we conducted experiments on disordered networks that were made out of foam sheets. When aged under uniform compression, the Poisson's ratio of these networks decreased. This previous work focused on the Poisson's ratio in the linear regime and its dependence on factors such as aging strain and aging time. By definition, the linear regime corresponds to small measuring strains where the Poisson's ratio does not change as a function of input strain. In the present paper, we focus on understanding the effects of aging in the nonlinear regime as can be probed by measuring the system's response at large strains. %We perform aging experiments on disordered networks where we measure the network response for a range of input strains. % and find interesting and complex behavior in the nonlinear regime.

\subsection*{Methods}
We consider quasi-two-dimensional  EVA(ethyl vinyl acetate)  foam sheets patterned in a disordered manner using a laser-cutter. Previously we have shown that the qualitative behavior is independent of the precise pattern~\cite{pashine2019directed}; here we select network-like patterns derived from simulations of jammed packings. The procedure for generating these networks is identical to that used in simulations, and is described in Sec.~\ref{simulations}. 
Since the dynamics are faster at higher temperatures, we store the networks in a convection oven or air sealed in a water bath at an ambient temperature of $50^{\circ} C$ while they age. This temperature is high enough that it significantly accelerates the aging process but not so high so as to damage the material. 

\begin{figure}
\includegraphics[scale=0.265]{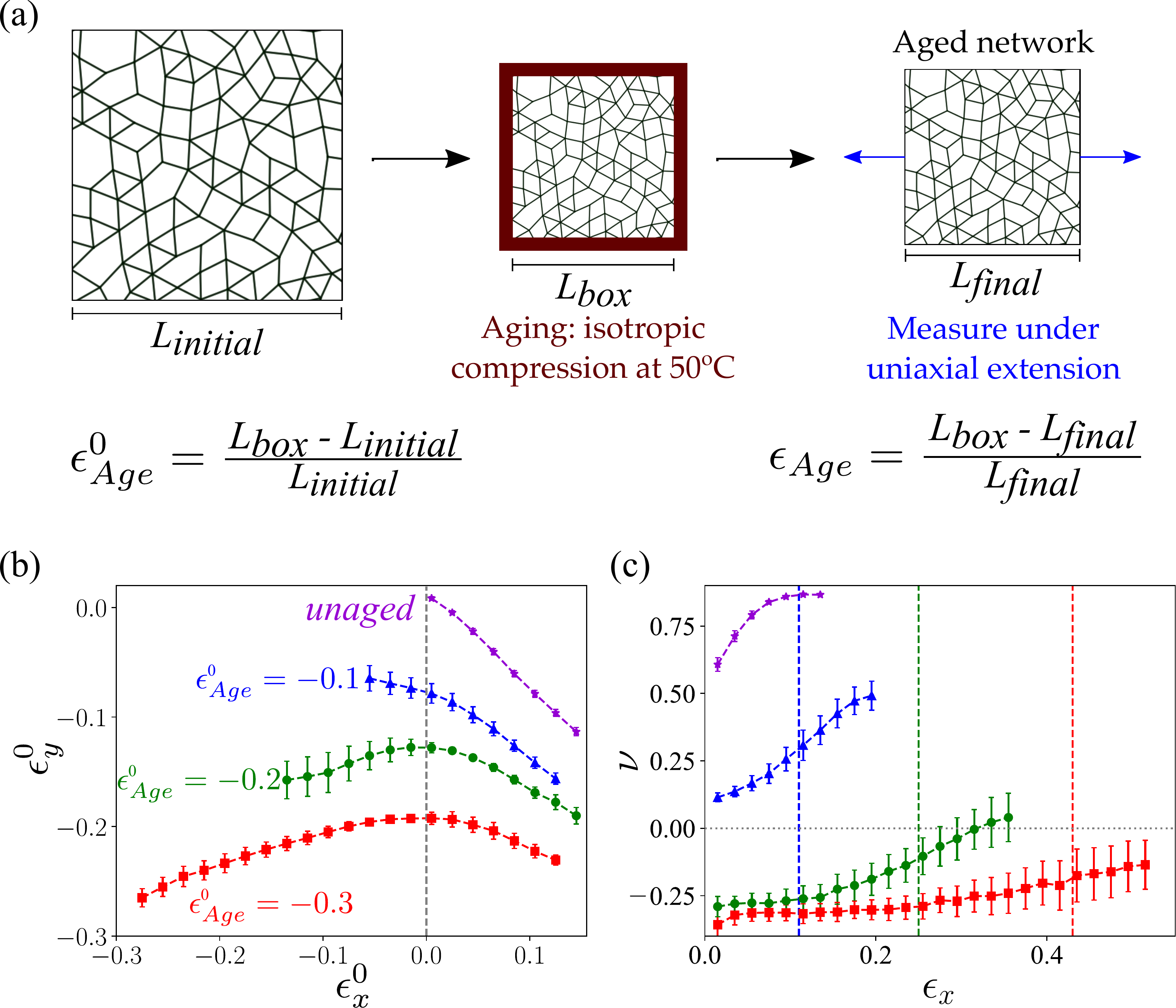}

\caption{Experiments of aging under isotropic compression. (a) A schematic of the experimental protocol. Our networks are aged under compression at an elevated temperature for a duration of one hour.  Once aged, these networks are removed from its confining box and immediately brought back to room temperature. The aged networks are measured under extensional uniaxial strain; $\epsilon^{0}$ is the strain measured with respect to the original unaged system, while $\epsilon$ is the strain of the network measured with respect to its aged size. (b) Aged networks are measured under uniaxial extension along the \textit{x} axis.  The vertical dashed line in grey corresponds to the unaged size, $L_{initial}$. Networks with data shown in blue (triangle), green (circle) and red (square) were aged at $\epsilon^{0}_{Age}= -0.1, -0.2$ and $-0.3$ respectively. (c) Poisson's ratio of aged networks vs strain for the same networks as in (b). Note that the strain $\epsilon_x$ is with respect to the aged system. Vertical dashed lines in blue, green and red mark $L_{initial}$ for data sets corresponding to $\epsilon^{0}_{Age}= -0.1, -0.2$ and $-0.3$ respectively.}
\label{fig:experiment}
\end{figure}

Our experimental protocol is shown as a schematic in Fig.~\ref{fig:experiment}(a). The network is confined in a box with an edge length of $L_{box}$, which is smaller than the length of the original network, $L_{initial}$. Our aging strain is therefore given by $\epsilon^{0}_{Age} = \frac{L_{box}-L_{initial}}{L_{initial}}$. The "0" superscript  in $\epsilon^{0}$ implies that the measurements are with respect to the initial unaged system, as opposed to $\epsilon$, which is measured with respect to the final aged system. %$$\epsilon^0$  therefore denotes the strain with respect to the mechanical equilibrium of the unaged system where there are no applied forces or constraints. 
Since the dimensions of the network change due to plastic deformations during aging, $\epsilon$ is usually different than $\epsilon^{0}$. 
Our aging protocol is to confine the network in a box and let it age for a duration of 1 hour at a temperature of $50^{\circ} C$. We then remove it from the confining box and measure the elastic response.

\subsection*{Results}
At the end of the aging process the networks are only slightly larger in size than the confining box itself. Our measurement protocol is to stretch the aged networks in the x-direction and measure the response in the y-direction. The hallmark of a nonlinear response is a nonlinear functional relation between x and y strains.

Figure~\ref{fig:experiment}(b) shows the response of such networks under uniaxial extension along $x$. When stretched along the $x$ axis, the network initially expands along the $y$ axis but then switches direction and starts to compress. This transition happens near $\epsilon^{0}_x = 0$, which corresponds to the size of the initial unaged network. The network has a non-monotonic response with a peak at $\epsilon^{0}_x = 0$.  This peak in the $\epsilon^0_y$ vs $\epsilon^0_x$ curve is a memory of the initial state.

Another way to characterize this response is to measure the Poisson's ratio ($\nu$) of aged networks. Poisson's ratio, $\nu$, is the negative
ratio of the transverse strain, $\epsilon_{y}$ to an imposed uniaxial strain $\epsilon_{x}$:
$\nu=-\frac{\epsilon_{y}}{\epsilon_{x}}$.  Figure~\ref{fig:experiment}(c) shows the Poisson's ratio of aged networks versus pulling strain, $\epsilon_x$. The vertical lines correspond to the initial size of the networks. %Taken together, these experiments suggest that the networks have a memory of the region in which they were trained $(\epsilon^{0} = 0$ to $\epsilon^{0} = \epsilon^{0}_{Age})$; in this region they have auxetic behavior.
These results show that aging alters the elastic response in a manner that imprints a memory of its initial state and by understanding this process, we can control the nonlinear strain-dependent elastic response.% These are the the main results of the paper.

\section{Modes of aging and numerical simulations}
\label{simulations}
We deliberately avoid the difficult task of capturing numerically the complex behavior of a real foam used in experiment. 
To understand the experimental results, we take a step back and consider the simplest possible model of a solid. The mechanical properties of solids have long been described within the harmonic approximation in terms of central-force spring networks. Each spring, indexed by $i$, has spring constant, $k_{i}$, and a rest length, $\ell_{i,0}$. The full aging behavior of our experimental systems depends on additional factors neglected here, such as the energy cost for varying the angle between adjacent bonds~\cite{reid2018auxetic} and the bending and buckling of bonds at large strains. In this approximation, however, the total elastic energy is then: 
\begin{equation}
U=\frac{1}{2}\sum_{i}k_{i}\left(\ell_{i}-\ell_{i,0}\right)^{2}.\label{eq:totalenergy}
\end{equation}
Here $\ell_{i}$ is the length of spring $i$.

 A system that is under imposed stress ages in a manner that will reduce its internal energy~\cite{Struik_1977_Polymer,Hodge_1995_Science}. It is clear that the system can accomplish this by varying the spring constants, $k_i$, and the equilibrium lengths, $\ell_{i,0}$. Again in the spirit of simplicity, we will disentangle these two effects by studying two models introduced in Ref.~\cite{pashine2019directed}. Despite the simplicity of our models we demonstrate striking qualitative similarity with the experiments.%In response to stresses on the bonds, the stiffness of each bond changes in the first model, introduced in Ref.~\cite{pashine2019directed}, while the microstructure geometry--determined by the equilibrium lengths--changes in the second model. 

 \textbf{k-model:} This model~\cite{pashine2019directed} captures the weakening of the bonds. %{\color{red} This isolates the effect that is emphasized by the relaxational protocol of the experiments, which we argue is dominated by local damage.} 
 We assume that the rate at which a bond becomes weaker depends on its energy so that it weakens both when compressed
or extended:
\begin{equation}
\partial_{t}k_{i}=-\gamma k_{i}\left(\ell_{i}-\ell_{i,0}\right)^{2}.\label{eq:k-model}
\end{equation}
Here $\gamma\equiv\frac{1}{\tau_{0}\left<\ell_{i,0}^{2}\right>}$, where
$\tau_{0}$ is a material-dependent relaxation time and $\left<\ell_{i,0}^{2}\right>$
corresponds to the average of the square lengths of the bonds before
aging. This model is similar in spirit to the design strategy in which bonds with large stresses are preferentially removed. \cite{goodrich2015principle}.

Generally, there are multiple ways by which a material's stiffness may  evolve under stress or strain. It is plausible for  material to get weaker as a result of the stress it experiences, and our model aims at understanding this effect. However, there are various examples where a material can also do the opposite, \textit{i.e.} get stronger due to stress. Work hardening is the strengthening of metals or polymers in response to strain~\cite{degarmo1997materials}. Bone may undergo remodelling in response to loads, by increasing the bone mass~\cite{WolffsLaw,huiskes2000effects}. Frictional contacts increase their contact area over time, strengthening their interface~\cite{bowden2001friction,dieterich1994direct,dillavou2018nonmonotonic}.

 \textbf{$\ell$-model:} This model isolates the effects of geometric plastic alterations.
 %, which are dominant in the quench protocol in experiments}.  
 We assume that the stress in a bond is reduced by changing its \emph{equilibrium length}. The rate of change
of the length depends on a bond's tension; they elongate
under tension and shorten under compression:

\begin{equation}
\partial_{t}\ell_{i,0}=\beta k_{i}\left(\ell_{i}-\ell_{i,0}\right).\label{eq:l-model}
\end{equation}
Here, $\beta\equiv\frac{1}{\tau_{0}\left<\ell_{i,0}\right>\left<k_{i}\right>}$,
$\left<\ell_{i,0}\right>$ is the average bond length and $\left<k_{i}\right>$
is the average spring constant.
When the rate at which a bond changes its length is much slower than the time to reach force balance, this model reduces to the Maxwell model for viscoelasticity~\cite{maxwell1867iv}. Each bond consists of a spring, which describes the rapid elastic behavior, in series with a dash-pot, which at
long times accounts for the change in rest lengths of the spring. Similar dynamics have also been used to account for junction remodelling in  epithelial cells~\cite{staddon2019mechanosensitive,cavanaugh2019rhoa}. Some geometrical effects not included in the $\ell$-model, such as the bending of bonds in the network, may also play a significant role.

%Note also that geometrical changes were responsible for the bulk of the response in those experiments. Thus, those experiments, in which physical networks were simply aged at a fixed strain, should be closer to the $\ell$-model than to the $k$-model, although 

We note that the two models can be expressed similarly and combined:

\begin{align}
\partial_{t}\sqrt{k_{i}} & =-\frac{\gamma}{4}\frac{\partial U}{\partial\sqrt{k_{i}}},\\
\partial_{t}\ell_{i,0} & =-\beta\frac{\partial U}{\partial\ell_{i,0}}.
\end{align}

We assume aging is much slower than the time to reach force balance. The microscopic parameters evolve by steepest descent
to minimize the energy at a rate proportional to the energy gradient.  

We study directed aging and response in each of these models numerically. To ensure that our networks are initially rigid, we derive the ensemble of central-force spring
networks from packings of soft spheres at force balance
under an external pressure \cite{Ohern,liu2010jamming,vanHecke_2015}.
The sphere centers define the locations of the nodes and overlapping
spheres are connected by springs. The equilibrium spring length is chosen
to be the distance between nodes, guaranteeing that in the absence
of deformation, the system is unstressed and at zero energy.
We characterize the connectivity of the network with the average coordination
number $Z=\frac{2N_{b}}{N}$, where $N_{b}$ is the number of bonds
and $N$ is the number of nodes. At the jamming transition, where the particles
just touch, $Z=Z_{c} = 2d$ is the smallest coordination number
needed to maintain rigidity in $d$ dimensions. 
We use networks that are above the jamming transition: $\Delta Z\equiv Z-Z_{c}>0$.

\subsection*{Aging under isotropic compression}
\subsubsection*{Aging protocol and evolution of the energy landscape in the $k$-model:}

We begin with an unstressed network and compress
it to the aging strain, $\epsilon^{0}_{Age}$. The strain is $\epsilon^{0}\equiv\frac{L-L_{initial}}{L_{initial}}$, where $L$ is the length of
the system while $L_{initial}$ is the length in the unstrained system.
Note that compression corresponds to a negative strain. To simulate quasistatic
dynamics we compress in small steps, minimizing the energy
with respect to the locations of the nodes at each step.  
Aging can be neglected during the measurement itself.

We measure the energy as a function of strain, $U\left(\epsilon\right)$, for isotropic
expansion and compression. In linear response, $U\left(\epsilon\right)=\frac{1}{2}VB\epsilon^{2}$,
where $V$ is the volume and $B$ is the bulk modulus. The initial
network has no internal stresses, so that the global energy minimum
is at zero strain where $U\left(\epsilon^{0}=0\right)=0$. Aging in the
$k$-model only changes the spring constants so the global
energy minimum remains at $\epsilon^{0}=0$. Since there is no difference in size between the unaged and aged system, $\epsilon^{0}_{Age} = \epsilon_{Age}$. After the system has been aged
at $\epsilon_{Age}$, we allow the system to return to its original volume and measure the behavior with respect to $\epsilon=0$ as origin. 

As noted above, aging reduces the energy at the strain at which the system has been aged. 
As shown in Fig. \ref{fig:constant_strain}(a), the energy
at the strain $\epsilon=\epsilon_{Age}$ is reduced after aging compared to the unaged material; the energy landscape has become very asymmetrical. 

Since the state at $\epsilon=0$
remains the global minimum with zero energy, there will be at least two states
with low energy $\epsilon=0$ and $\epsilon=\epsilon_{Age}$.  If $\epsilon_{Age}$ is small, the low-energy states encompass the entire range $\left[0,\epsilon_{Age}\right]$; however, if $\epsilon_{Age}$ is large, aging can result in two local energy minima separated by an energy barrier.

The low-energy state at $\epsilon_{Age}$, is a memory in the landscape of the conditions under which the system was aged. By measuring the elastic properties, the aging strain can be read
out.

\subsubsection*{Results for Poisson's ratio of $k$-model aged under isotropic strain:} 
Within linear response for an isotropic material, Poisson's ratio, $\nu$ is a monotonic function of the ratio
of the shear modulus, $G$, to the bulk modulus, $B$: $\nu=\frac{d-2G/B}{d\left(d-1\right)+2G/B}$; where $d$ is the spatial dimension. At larger strains, $\nu$ generically depends on the magnitude of $\epsilon$.  We measure $\nu$ by compressing (or expanding) the system uniaxially, while allowing the system to relax in the transverse directions. If the system is isotropic then the transverse strain, $\epsilon_{r}$ is the same in all transverse directions.

%In this paper, we are interested in creating auxetic (\textit{i.e.}, $\nu<0$) materials.These are not ubiquitous in nature and are thought to have interesting applications~\cite{greaves2011poisson,ren2018auxetic}. We show that, simply by aging, a system can be transformed into an auxetic material.  This demonstrates the power of the directed-aging protocol.

\begin{figure}
\begin{centering}
\includegraphics[scale=0.58]{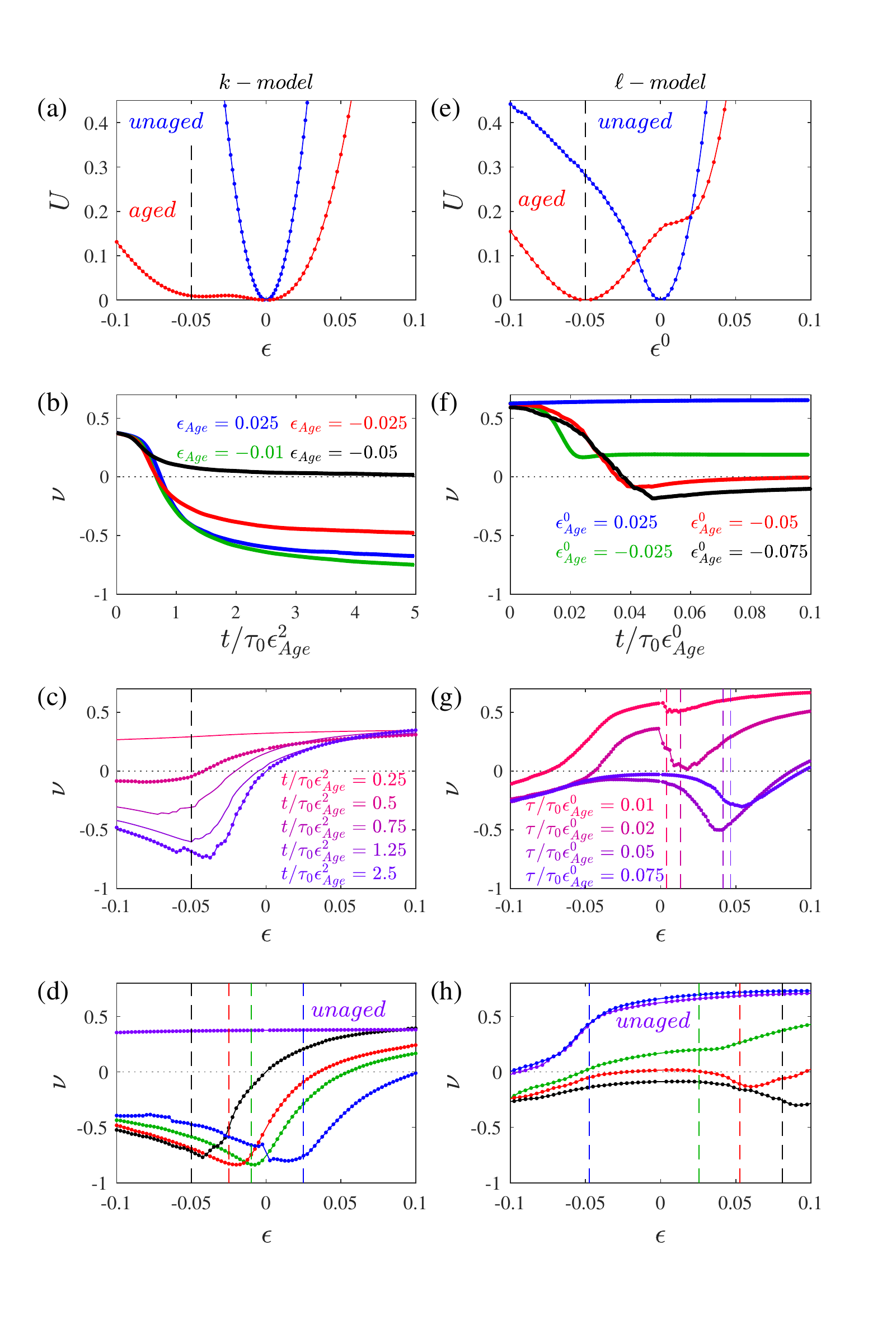} 
\par\end{centering}
\caption{ Left: the $k$-model at $\Delta Z\approx1.51$. Right: $\ell$-model
at $\Delta Z\approx0.53$. (a), (e) The compression energy versus strain for an unaged system and a system aged under $5\%$
compression (dashed line). In the $k$-model
the global minimum remains at zero strain, since bond lengths do not
change while in the $\ell$-model the global minimum 
shifts. The difference in the unaged energy versus strain between the two models is caused by the two different values of $\Delta Z$ (picked to allow comparison of the two models at the same range of strains). (b) and (f) The Poisson's ratio within linear response versus time.
In the $\ell$-model the curve is non-monotonic and has
a minimum. (c) and (g) The evolution of the Poisson's ratio as a function
of strain. In the $k$-model, the local minimum occurs near the aging
strain, $\epsilon_{Age}=-0.05$ (vertical dashed line). In the $\ell$-model
the minimum corresponds to the strain needed to undo the volume change that
occurred during aging (dashed lines). (d) and (h) The
Poisson's ratio versus strain. For the $k$-model 
these are measured at a constant $\epsilon_{Age}^{2}t$.
The aging strain is denoted by vertical dashed lines.
In the $\ell$-model, the curves are at a constant $t\epsilon_{Age}$ and the dashed lines
denote the strain corresponding to the unaged system at asymptotic
times, $-\epsilon_{Age}/\left(1+\epsilon_{Age}\right)$. Note that in
(e) the strain is measured with respect to the unaged system while
for (g) and (h) it is measured with respect to the new global minimum,
which depends on the aging time and $\epsilon_{Age}$. \label{fig:constant_strain}}
\end{figure}

Fig.~\ref{fig:constant_strain}(b) shows that the Poisson's ratio in the linear response regime, $\nu(\epsilon\rightarrow0)$  decreases as the network ages
and ultimately may become negative. At long times, $\nu$ monotonically decreases with $\left|\epsilon_{Age}\right|$. This is consistent with the system aging in a directed manner~\cite{pashine2019directed}; that is, aging under compression lowers the bulk modulus more than
it lowers the shear modulus. 

\begin{figure}

\includegraphics[scale=0.5]{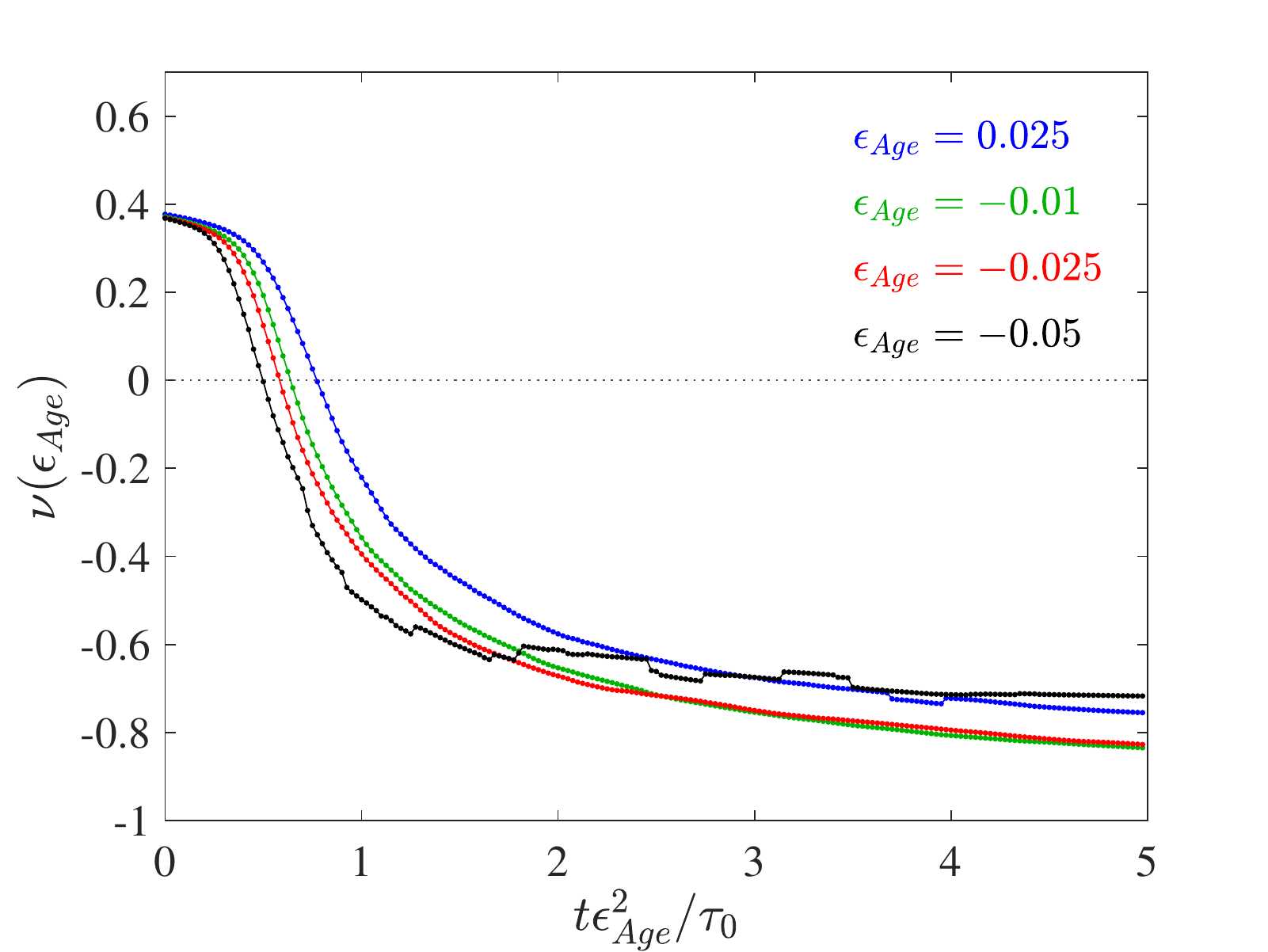}

\caption{In the $k$-model, the Poisson's ratio at $\epsilon=\epsilon_{Age}$ as a function of scaled time approximately collapses.\label{fig:collapse_kmodel} }

\end{figure}

The relaxation rate has a characteristic dependence on the imposed
strain. The right-hand side of Eq.~\ref{eq:k-model} scales as as $\epsilon_{Age}^{2}$, the
energy of the system. To compare different aging strains, we rescale time in Fig.~\ref{fig:constant_strain}(b) as $t\epsilon_{Age}^{2}$. In Fig.~\ref{fig:collapse_kmodel} we show that $\nu$ at $\epsilon=\epsilon_{Age}$ approximately collapses as a function of $t\epsilon_{Age}^{2}$.

The nonlinearities in the energy landscape suggest that $\nu$ may
have interesting dependence on the strain. We show
how $\nu\left(\epsilon\right)$ evolves with time in Fig.~\ref{fig:constant_strain}(c). For the unaged systems, $\nu$ depends very weakly
on strain even up to 10\% strain. %This weak strain dependence is due to the large connectivity in the system, which results in the system having a linear regime over a broad range of strains. 
As the system
ages, $\nu\left(\epsilon\right)$ is lowered, especially near $\epsilon_{Age}$.
It later develops a minimum near the aging strain, that deepens
over time. 

In Figure~\ref{fig:constant_strain}(d)  we plot $\nu\left(\epsilon\right)$
for different values of $\epsilon_{Age}$, at the same scaled time $t\epsilon_{Age}^{2}$.
In all the curves, a minimum appears in $\nu\left(\epsilon\right)$
close to $\epsilon =\epsilon_{Age}$ (vertical dashed lines). Thus, aging produces
a memory of the strain at which it was prepared, allowing the aging
strain to be read out from the minimum of the non-linear Poisson's
ratio. %This is similar to the memory in Fig.\ref{fig:experiment}(a) that we see in our experiments.

Aging in the \textit{non-linear} regime allows exotic behavior to
be trained into a system. For example, a system can have a positive
$\nu$ within linear response but a large negative $\nu$ for larger
unixial compression (see the curve for the aging strain of $\epsilon_{Age}=-0.05$
in Fig.~\ref{fig:constant_strain}(d)). Therefore, compressing uniaxialy
leads initially to a transverse expansion, which is then reversed
to transverse compression at larger values of strain.

\subsubsection*{Aging protocol and evolution of the energy landscape in the $\ell$-model:}
In the $\ell$-model, the rest lengths of bonds, $\ell_{i,0}$, evolve as the system ages under an imposed strain to reduce the stresses and the elastic energy. We compress the system to the target strain, $\epsilon^{0}_{Age}$, and then
allow it to age. Since the rest lengths have evolved, the global energy minimum is no longer at zero strain. Since the energy minimum shifts, we measure the elastic properties with respect to the \textit{new} (aged) global minimum $(\epsilon = 0)$. 

At long times, all the stresses decay
to zero and the global energy minimum shifts to $\epsilon^{0}_{Age}$ as shown in Fig.~\ref{fig:constant_strain}(e). At intermediate time, before it reaches that point, the minimum lies between $\epsilon^{0}=0$ and $\epsilon^{0}=\epsilon^{0}_{Age}$. To find the position of the minimum, we minimize the energy with respect to the node locations, as well as the width and height
of the box. 

%Note that the network shown for the \textit{$k$-model} in Fig.~\ref{fig:constant_strain}a  has a larger coordination number $Z$ than the one used for the $\ell$-model in Fig.~\ref{fig:constant_strain}e. While this choice is not essential, it allows a comparison of the different effects at the same range of strains between the two models.  This is the cause of the difference between the two models for the unaged energy versus strain.  

Besides the shift of the energy minimum, there are additional changes
to the energy versus strain curves apparent in Fig.~\ref{fig:constant_strain}(e).  First, we note that the curvature
at the new global minimum, which determines the bulk modulus, $B$, is greatly reduced. This
lowers the Poisson's ratio, as we will discuss in detail below. Another
feature is the kink in the nonlinear regime near $\epsilon^{0}=0$, corresponding
to unstressed state of the unaged system ($\epsilon^{0}=0$ corresponds
to a strain of $\epsilon=-\frac{\epsilon^{0}_{Age}}{1+\epsilon^{0}_{Age}}$ measured
with respect to the aged system). 
This is a signature of the history of the strains at which the
system was aged.

\subsubsection*{Results for Poisson's ratio of $\ell$-model aged under isotropic strain:}
We first consider the evolution of the Poisson's ratio within linear response, $\nu(\epsilon\rightarrow0)$, as shown in Fig.~\ref{fig:constant_strain}(f).
Aging under constant compression results in the linear
Poisson's ratio decreasing
with time. The decay is non-monotonic,  with a
local minimum at intermediate times.  At large times it ceases
to evolve and its asymptotic value depends on $\epsilon^{0}_{Age}$.  At small values of $\left|\epsilon^{0}_{Age}\right|$, $\nu$ remains positive.  For systems aged at larger $\left|\epsilon^{0}_{Age}\right|$, however, $\nu$ decreases further and ultimately becomes negative for sufficiently large aging strain $\left|\epsilon^{0}_{Age}\right|$.  This is similar to
the behavior seen in experiments on aging networks~\cite{pashine2019directed}. 

By contrast to the behavior under compression, aging under a small constant expansion  (blue curve) increases the linear-response Poisson's ratio, $\nu$ only slightly. 
We believe that this is a general feature
of elastic systems and is due to mechanical instabilities that occur under compression but not under expansion.
These instabilities reduce the stiffness to compression significantly, thus lowering $\nu$. Instabilities have also been shown to give rise to materials with negative Poisson's ratio in periodic structures~\cite{bertoldi2010negative}.

The nonlinear Poisson's ratio versus strain is shown for different
times in Fig.~\ref{fig:constant_strain}(g). These curves have
a minimum at $\epsilon>0$ (expansion), and the value of strain grows
with time.  The dashed lines in Fig.~\ref{fig:constant_strain}(g)
showing the applied strains approximately match the location of the minima. This suggests that the minimum is associated with the strain needed to return to the same volume
as the unaged system. 

We next consider the nonlinear behavior of the $\ell$-model at long aging times.
 Figure~\ref{fig:constant_strain}(h) shows the Poisson's ratio versus strain for the unaged system and
for systems aged at different values of $\epsilon^{0}_{Age}$. 
For the unaged system, $\nu\left(\epsilon\right)$ decreases when the system is compressed. At positive strains the aged systems develop a local minimum that is a memory of the strain at which
they were aged. Unlike in the $k$-model, where the memory occurs at
$\epsilon=\epsilon_{Age}=\epsilon^{0}_{Age}$, the minimum in this case occurs near the strain which
corresponds to that of the unaged system ($\epsilon^{0}=0$). Aging under
compression reduces the volume and to return to the initial volume
the system must be strained by $-\frac{\epsilon^{0}_{Age}}{1+\epsilon^{0}_{Age}}$.
This is indicated by dashed lines in Fig.~\ref{fig:constant_strain}(h). The memory observed in these simulations is similar to the experiments discussed in Sec.~\ref{experiments1}, where we see a memory of initial system in the non-linear elastic response.

We believe, that the microscopic mechanical instabilities, such as buckling of bonds, play an important role in encoding these memories. Such instabilities occur when a compressive strain is applied to the system. Expanding a system that was aged under compression causes some of these instabilities to be ``undone''. However, once this system is stretched to its initial size, there are no more of these instabilities left. This results in the unusual nonlinear response near $\epsilon^{0}=0$.

\subsubsection*{Discussion of simulation results for aging under compression}
%\textcolor{blue}{The experimental data provides strong evidence that the models we use in simulations capture effects that can be realized in real materials.} 

%Our simulation results show two distinct memories that arise in our systems: a memory of the aging strain in the $k$-model and a memory of the initial state in the $\ell$-model. Since these memories define the elastic behavior of the system, they can be used to modify and control the non-linear elastic response of a material.

Our simulations show that the two models evolve the system in two distinct ways. In the $k$-model, the equilibrium state remains unchanged but the system gains a memory of the aging strain. On the other hand, the energy minimum in the $\ell$-model moves to the aging strain, and the system develops a memory of its initial state.

The similarity in the experimental results from Sec.~\ref{experiments1} and the simulation results for the $\ell$-model imply that our experiments are dominated by the effects of geometry change in the aged networks. Our simulation results for $k$-model inspired us to perform further experiments that highlight the effect of material weakening. We present these results in Sec.~\ref{experiments2}.

%We see both of these memories in the experiments. 

%Taken together our experiments and simulations on aging under isotropic compression show that physical networks can be manipulated by the aging protocol and that memories dictate the behavior of the system deep in the non-linear regime.

\subsection*{Aging under shear}

\begin{figure}
\includegraphics[scale=0.58]{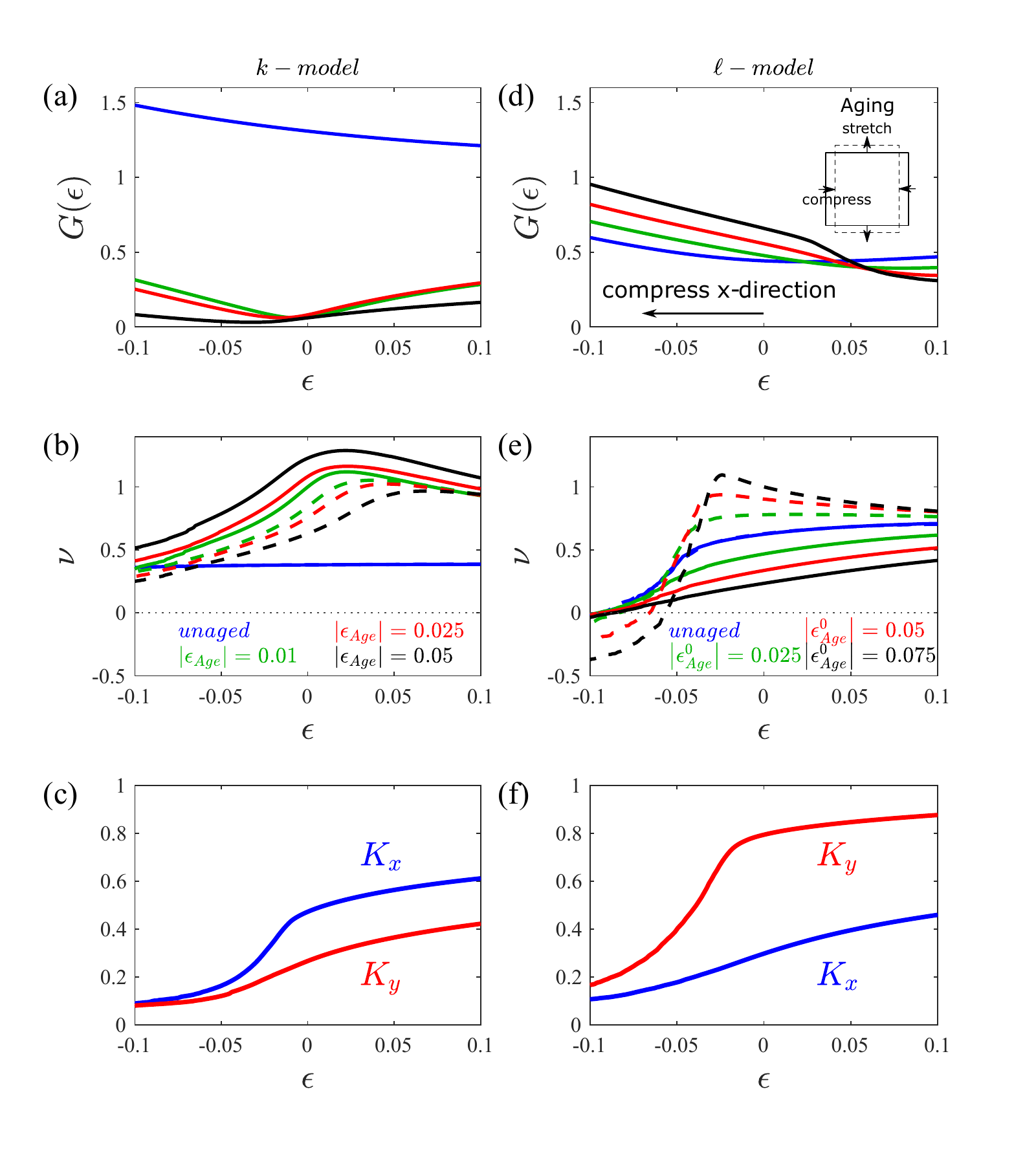}

\caption{Aging under shear. Left: $k$-model. Right: $\ell$-model. The system
is aged by compressing along the x-axis and stretching along the y-axis.
(a) and (d) the shear stiffness %defined by the energy per unit strain squared, 
versus measuring strain, $\epsilon$. (b) and (e)
the Poisson's ratio versus strain. The full line denotes,
$\nu_{x}$, the Poisson's ratio when the system is strained along the
x-axis. The dashed line denotes, $\nu_{y}$, measured by straining
along the y-axis.(c) and (f) The stiffness to uniaxial deformation, $K_x$ and $K_y$ along the   x-axis and y-axis. Note that after aging, $K_x>K_y$ in $k$-model while $K_x<K_y$ in the $\ell$-model. \label{fig:shear}}

\end{figure}
To test the generality of the nonlinear effects of aging, we consider  systems that are aged under a constant shear strain rather than under compression. We demonstrate that the ensuing change in elasticity is more subtle and does not necessarily follow the intuition gleaned from the case of aging under compression.

We shear the elastic networks by compressing along
the x-axis by $\epsilon^{0}_{Age}$ and extended along the y-axis by $-\frac{\epsilon^{0}_{Age}}{1+\epsilon^{0}_{Age}}$
to preserve volume. As before, we characterize the elastic behavior
by measuring the Poisson's ratio. However, aging under shear evolves
the system to be anisotropic so that $\nu$ depends
on the direction of applied strain. We focus on the Poisson's ratio
measured by straining either along the x-axis ($\nu_{x}$), or y-axis ($\nu_{y}$), while the
system is allowed to relax in the transverse direction.\\

\subsubsection*{Results for Poisson's ratio of $k$-model aged under shear strain:}
In Fig. \ref{fig:shear}(a) we show the stiffness
to shear, measured from the elastic energy per unit strain squared,
$G\left(\epsilon\right)=2U_{G}/V\epsilon^{2}$. In the limit of $\epsilon\rightarrow0$
this corresponds to the linear-response shear modulus. As expected,
aging under shear lowers $G\left(\epsilon\right)$. The strain dependence
of $G\left(\epsilon\right)$ shows additional features. 
In particular $G\left(\epsilon\right)$ has a minimum at a value of strain that depends
on the aging strain. The minima occur at a strain that is slightly
different than $\epsilon_{Age}$ but appears to be proportional to $\epsilon_{Age}$. Note that $\epsilon_{Age} = \epsilon^{0}_{Age}$.
These minima also encode memories of how the system was prepared.

Figure~\ref{fig:shear}(b) shows the Poisson's ratio as a function of
strain for different aging strains. The full line denotes,
$\nu_{x}$, the Poisson's ratio when the system is strained along the
x-axis while the dashed line denotes $\nu_{y}$, measured by straining
along the y-axis. As expected, aging under
shear decreases the associated shear modulus, and therefore increases the Poisson's
ratio in that direction. We also find that for a given value of strain $\nu_{x}>\nu_{y}$. 
The difference between the Poisson's ratio in these two directions grows with $\left|\epsilon_{Age}\right|$. Both decrease for compression, and are peaked at a given strain for expansion. The strain of the peak in $\nu_{x}$ depends weakly on the aging strain, while the strain of the  peak in $\nu_{y}$ grows with the aging strain.

The result $\nu_{x}>\nu_{y}$ can be understood as follows in linear response, where
 $\nu_{x}=\frac{B+G}{4K_{y}}$, and $\nu_{y}=\frac{B+G}{4K_{x}}$,
where $K_{y}$ and $K_{x}$ are the stiffnesses to uniaxial compressions
along the x-axis and y-axis respectively. The inequality $\nu_{x}>\nu_{y}$ implies $K_{x}>K_{y}$, which is shown in Fig. \ref{fig:shear}(c). 

The difference between $K_{x}$ and $K_{y}$ is the
result of the asymmetry between compression and expansion in the non-linear
regime in which the system is aged. We find that the stiffness to compression
in the nonlinear regime is smaller than for expansion. Therefore,
the stresses along the y-axis are larger, and as a result $K_{y}$
ages faster and becomes weaker than $K_{x}$. Thus, aging at 
shear strains in the nonlinear regime can affect different linear response moduli
differently.

\subsubsection*{Results for Poisson's ratio of $\ell$-model aged under shear strain:} 
Fig. \ref{fig:shear}(d) shows the stiffness
to shear as a function of the measuring strain, $G\left(\epsilon\right)$.
For negative strains (\textit{i.e.} in the direction  which the system was aged)
the stiffness grows, while for large positive strains the stiffness
decreases. This is different from the case of aging under compression,
where stiffness to compression decreases. Similar behavior has been
reported~\cite{majumdar2018mechanical} in sheared gels. There the increase in stiffness
was attributed to bonds aligning in a preferential direction defined
by the shear deformation. Prior to shearing, the system was isotropic
and each bond angle is equally probable. The shear deformation, in
our case, tends to align the bonds along the y-axis. This increases
the uniaxial stiffness in the y-direction, $K_{y}$, and reduces the
stiffness in the x-direction, $K_{x}$ (see Fig. \ref{fig:shear}(f)). This is consistent with experiments~\cite{pashine2019directed} and explains why in linear response $\nu_{y}>\nu_{x}$, as shown in Fig. \ref{fig:shear}(e).

The nonlinear behavior of $\nu_{x}$ and $\nu_{y}$ has a different
dependence on the shear strain, as shown in Fig. \ref{fig:shear}(e).
The slope of $\nu_{x}$ and $\nu_{y}$ have opposite signs. Interestingly,
$\nu_{y}$ has a peak at negative values of strain and then falls
steeply. We believe, that as in the case for compression, the sharp
drop in $\nu_{y}$ is associated with instabilities.
The threshold value of strain for these instabilities decreases with aging strain. 

In summary, in both models aging under shear gives rise to a change in
the stiffness to shear. The direction in which these evolve are
not always obvious, and also result in nontrivial nonlinear behavior.

In real systems, local bond stiffnesses can be weakened (\textit{$k$-model}) and local geometry can change ($\ell$-model). The differences between the behavior of the \textit{$k$-model} and  $\ell$-model could allow one to determine which of the two aging scenarios that they embody is dominant for a given system at a given strain. In the case of the $k$-model we find that $\nu_{x}>\nu_{y}$ while in the $\ell$-model $\nu_{x}<\nu_{y}$. In addition, the slope of $\nu_{y}$ at $\epsilon=0$ has the opposite sign for the two models.

\subsubsection*{Discussion of simulation results for aging under shear}

Ref.~\cite{pashine2019directed} considered the effect of aging under a constant shear strain in experiments on foam networks, similar to those used here. There the system was not cycled to minimize the permanent deformation, and the initially square network became rectangular. We follow the same notation used for simulation where shear is performed by compressing along the x-axis while extending along the y-axis.

In the unaged system the Poisson's ratio is virtually independent of direction, and $\nu^{unaged}_x\approx\nu^{unaged}_y$. After aging the response becomes highly non-isotropic, with $\nu_y>\nu^{unaged}>\nu_x$ ~\cite{pashine2019directed}. This is consistent with the scenario of aging in the $\ell$-model, shown in Fig. \ref{fig:shear}(e). Though the comparison is only qualitative, it demonstrates that these models can capture complex behavior that arises in real materials. It opens the door to quantitative studies in the nonlinear regime and over a broader range of parameters.

\section{Additional experiments motivated by simulations}
\label{experiments2}
The distinct results seen numerically for the $k$-model and $\ell$-model suggest that it would be enlightening to develop experimental protocols that accentuate either material weakening (isolated in the $k$-model) or geometrical changes (isolated in the $\ell$-model). Note that in Ref.~\cite{pashine2019directed} we were able to test the effects of purely geometrical changes by taking an image of aged networks and laser-cutting a new network with the same geometry. By this protocol, the effects of aging on spring constants captured by the $k$-model are eliminated. Here we introduce an experimental protocol that accentuates the physics captured by the $k$-model.

\emph{Relaxational protocol (stiffness-dominated aging):}
We start with letting the laser cut, EVA foam networks age under isotropic compression at an elevated temperature of $50^{\circ} C$ for an hour, exactly like our previous experiments in Sec.~\ref{experiments1}. However, instead of bringing them to room temperature immediately afterwards, we let them relax at the same ambient temperature of $50^{\circ} C$ for 1 hour while unconfined. As it relaxes, the network returns partway to its original size and shape.  This reverses some of the geometric change that occurred during the aging process. The elevated temperature causes  more rapid relaxation than at room temperature. After an hour, we bring the network back to room temperature. At this point, the relaxation dynamics has slowed down considerably and the network does not evolve much further. This is the limit where the system has had a chance to relax after aging under stress and has regained most of its original geometry. The major contributor to the change in material properties in such a system would be the microscopic `damage' that occurs in the material. Thus, results from this protocol should be better described by the $k$-model than the $\ell$-model.

\begin{figure}
\includegraphics[scale=0.265]{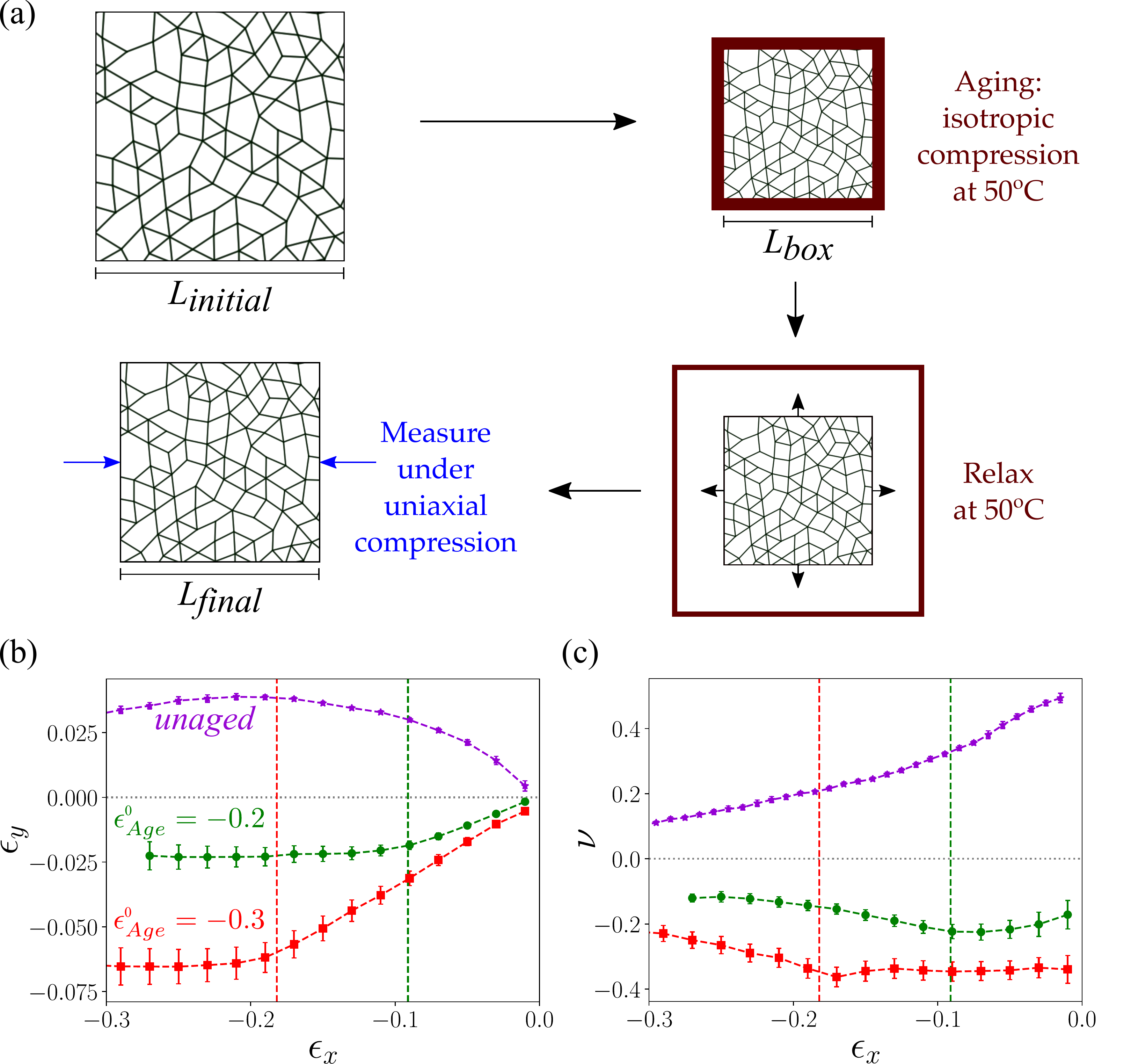}

\caption{Experiments of stiffness-dominated aging under isotropic compression. (a) A schematic of the experimental protocol. The networks are allowed to relax after aging under compression. The aged networks are then measured under compressive uniaxial strain. (b) Networks are compressed uniaxially along \textit{x} axis and their response is measured along \textit{y} axis. $\epsilon$ is the strain of the network measured with respect to its aged size. Purple (star) data corresponds to unaged networks. Green (circle) and red (square) data are from networks aged at two different strains. Vertical dashed lines in red and green correspond to the physical size to which they were compressed during aging, $L_{box}$. (c) The Poisson's ratio vs strain for the same set of experiments as (b).}
\label{fig:experiment2}
\end{figure}

Fig.~\ref{fig:experiment2}(b) shows $\epsilon_y$, the measured strain of the networks along the $y$ axis, as a function of $\epsilon_x$, the applied strain along $x$ axis. We focus on compressive strains ($\epsilon_x<0$) since that is the regime the system encounters during aging. Both strains are measured with respect to the aged system which has a length $L_{final}$. The purple curve shows that even the unaged system is nonlinear in the large strain regime we consider. We note that the negative slope near $\epsilon_x=0$ implies that the Poisson's ratio, $\nu=-\epsilon_y/\epsilon_x$ is positive.

The aged systems are shown in green (circle) and red (square) for aging strains of  $\epsilon^{0}_{Age} = -0.2$ and $-0.3$ respectively. Since $L_{final}<L_{initial}$, they correspond respectively to $\epsilon_{Age} \approx -0.09$ and $-0.18$ (represented by vertical dashed lines). The first feature we point out is that the curves have a positive slope near $\epsilon_x=0$, which indicates that the Poisson's ratio is negative. This is consistent with previous findings~\cite{pashine2019directed,lakes_r_Science_1987}. The interesting feature here is the change in slope as the system approaches the aging strain (vertical dashed lines). This is a memory of the aging strain visible in the non-linear response of the system. This shows that simply by measuring the strain response in the non-linear regime, we are able to deduce the aging strain of the material.

Alternatively, this effect can be seen in Fig.~\ref{fig:experiment2}(c) as a dip in Poisson's ratio near the aging strain.
These results are qualitatively very similar to the simulation results for the $k$-model under compression, discussed in Sec.~\ref{simulations}. A memory of aging strain was seen in  Fig.~\ref{fig:constant_strain}(d), where the Poisson's ratio has a minimum very close to the aging strain.
%Poisson's ratio as a function of applied strain with the minimum occurring very close to the aging strain. 
The similarity in these two results show that we can perform experiments in a way that accentuates the `microscopic damage' aspect of aging as opposed to the macroscopic geometry change. We also learn that our relatively simple simulations of aging under the $k$-model capture the real effects seen in experiments.

Together with the experiments presented in Sec.~\ref{experiments1}, we show that the memories seen in the two models of aging exist in real physical systems. Not only do these memories reveal the aging history of these networks, but they also provide a powerful tool for controlling the non-linear elastic properties of the system.

\section{Inherent nonlinear tunability of the Poisson's ratio} 
We have shown that aging at finite isotropic or shear
strains, $\epsilon_{Age}$, allows the Poisson's ratio to be manipulated in the nonlinear elastic regime as well as in the linear regime.
The ability to manipulate the non-linear behavior does not only depend
on the aging dynamics but is an inherent property of how forces are
transmitted in the network. In the linear regime, the pruning of bond $i$ leads to a change in the bulk modulus, $\Delta B_i$ that is uncorrelated with the change in the shear modulus, $\Delta G_i$~\cite{goodrich2015principle,hexner2018linking}. It is this property  --  the independence of bond-level response  --  that allows the linear-regime Poisson's ratio to be tuned so successfully by pruning. 

To quantify the inherent tunability of the Poisson's ratio in the \emph{nonlinear} regime,
we consider the contribution of a single bond to the modulus~\cite{goodrich2015principle}, which we denote for compression $B_{i}\left(\epsilon\right)=k_{i}(\delta x_{i}^{B})^{2}/\epsilon^{2}$, where 
the extension $\delta x_{i}^{B}$ depends on the amplitude of the
imposed strain. Essentially, this is the energy in a single bond per square unit strain; in the linear regime, it reduces to the contribution of bond $i$ to the linear bulk modulus.
If $B_{i}$ is a constant independent of the amplitude of $\epsilon$ for all $i$, then the elastic behavior does not depend on $\epsilon_{Age}$; the aging strain only changes the aging rate. The nonlinear correlations are characterized by comparing the correlations of $B_{i}\left(\epsilon\right)$
with their corresponding values in linear response. We use the Pearson correlation
function of two random variables, which is defined as: $C\left(y,x\right)=\left[\left\langle xy\right\rangle -\left\langle x\right\rangle \left\langle y\right\rangle \right]/\sigma_{x}\sigma_{y}$, where $\sigma$ is the standard deviation.

\begin{figure}
\includegraphics[scale=0.6]{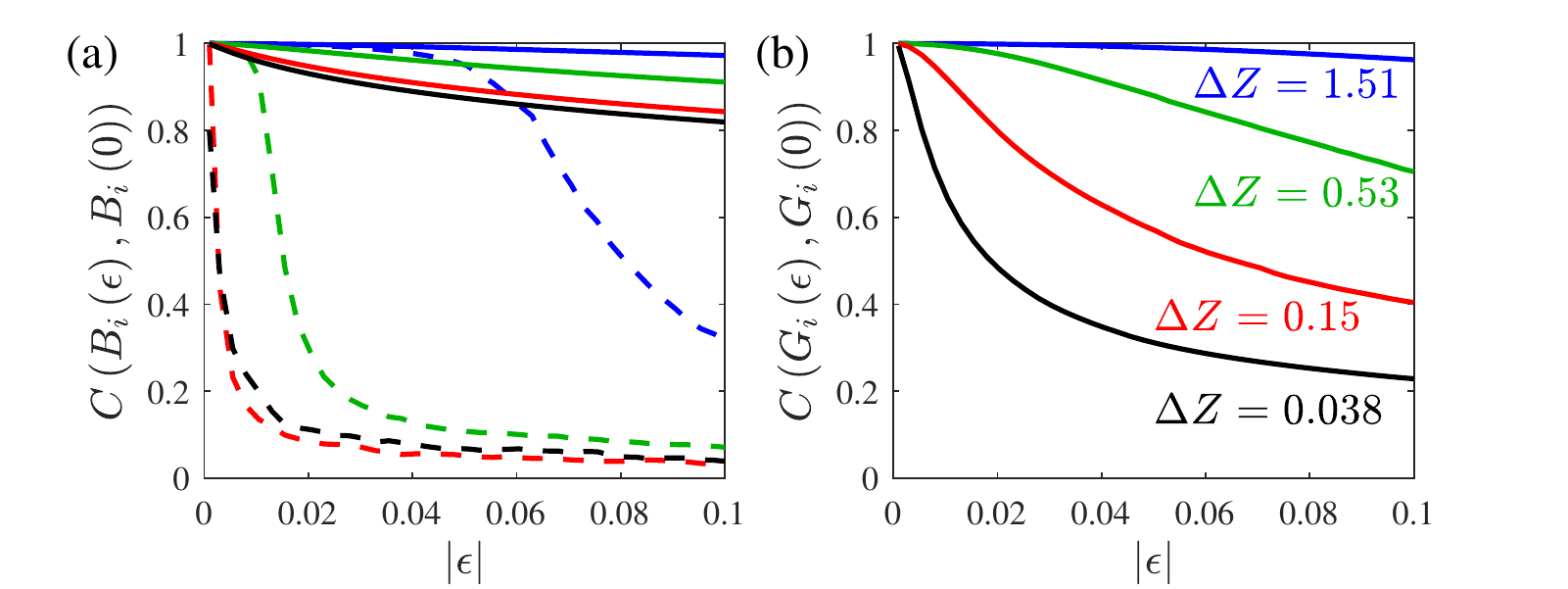}

\caption{Correlation between the elastic response at different strains for the unaged system. (a)
The Pearson correlation function of $B_{i}\left(\epsilon\right)$
with $B_{i}\left(0\right)$. Expansion is denoted by the full line 
and compression is denoted by a dashed line. Correlations for compression
decay more rapidly than for expansion.  The different colors are for different values of $\Delta Z$ as shown in legend of (b). %presumably due to buckling.
(b) Correlations of $G_{i}\left(\epsilon\right)$ and $G_{i}\left(0\right)$
decay faster for lower coordination number, $\Delta Z$. Since the unaged system is on average isotropic $G_i$ correlations are equal for positive and negative strains.  \label{fig:BiBi}}
\end{figure}

Fig.~\ref{fig:BiBi}(a) shows $C\left(B_{i}\left(\epsilon\right),B_{i}\left(0\right)\right)$
versus strain for both compression and expansion. Depending on the
coordination number, $\Delta Z$, correlations under compression decay
differently from those under expansion: the correlations are similar
below a threshold value of $\epsilon$, but then drop much more rapidly
in the case of compression. This threshold, which appears to vanish in the
limit of $\Delta Z\rightarrow0$, signals the breakdown
of linear response at the level of a single bond. Correlations under
expansion remain significant even for small $\Delta Z$.

The ability to tune the non-linear response to shear strain is characterized
by $C\left(G_{i}\left(\epsilon\right),G_{i}\left(0\right)\right)$,
as shown in Fig.~\ref{fig:BiBi}(b). Here, $G_{i}\left(\epsilon\right)=k_{i}(\delta x_{i}^{G})^{2}/\epsilon^{2}$
is measured at a shear strain of $\epsilon$. The correlations decay
faster when $\Delta Z\rightarrow0$. This implies that the system
is more  tunable in the non-linear regime near $\Delta Z\rightarrow0$. 

We can also characterize the ability in the nonlinear regime to tune the bulk and shear moduli independently. Within linear response in two dimensions, the correlation, $C\left(B_{i}\left(\epsilon\right),G_{i}\left(\epsilon\right)\right)$, was found to be small, $\approx0.17$~\cite{goodrich2015principle}. At a finite compressive strain we find that this correlation becomes even smaller. Under expansion, the correlations increase but remain below 0.33. 

We note that these correlations provide a measure of the inherent ability to tune a given modulus in the nonlinear regime. They do not depend on the protocol by which the system evolves under aging. They depend only on the nonlinear elasticity and could be used to identify interactions or geometries that are particularly amenable to manipulation by aging.

Our experimental results raise many interesting questions about the nature of aging in these systems. For example, the time dependence of aging and the effects of multiple aging cycles are not well understood. These results open up various avenues for further research that would allow us to understand the limits to which we can train a material.

\section{Discussion}
%In conclusion, aging evolves a mechanical system so that the elastic properties of the system change depending on the imposed deformation and strain. This occurs through changes in the microscopic parameters as the system minimizes its elastic energy. 

Minimization of a cost function is a common design method for achieving
certain target properties, such as auxetic behavior. While this is effective on a computer at sufficiently small system sizes, minimization is usually not possible to implement in the laboratory and is not scalable to arbitrarily large systems. Directed aging evolves a mechanical system so that the elastic properties of the system change depending on the imposed deformation
and strain. In contrast to cost-function minimization by computer, directed aging can be performed on systems of any size both on the computer and in the laboratory~\cite{pashine2019directed}.  Here we have shown that directed aging is highly effective at tuning the Poisson's ratio even in the nonlinear elastic regime. This result is important to many applications, including impact mitigation \cite{alderson1994auxetic,sanami2014auxetic} and filtration \cite{alderson2000auxetic,alderson2001auxetic}. %In the \textit{$k$-model} this is robust, and by aging at different strains the non-linear response can be controlled. In the $\ell$-model, aging under compression allows auxetic behavior that depends on the aging strain.

We find that directed aging profoundly affects
the nonlinear behavior in a way that can be very different from how
it changes the linear response. In some cases the Poisson's ratio
can be manipulated so that it changes sign as a function of strain.
This could allow a designer to manipulate the energy landscape and
create material with desired properties that vary as a function of
the amplitude of deformation. More complex energy landscapes could
clearly be achieved by varying the strain as the system ages. 

Remarkably, the behavior in our minimal models is very similar to the aging experiments we conduct in foam samples. In both cases there is a memory of the strain to which they are aged, and it can be read out from the minimum of the Poisson's ratio. We note however, our models neglect effects such as the energy cost for varying the angle between adjacent bonds~\cite{reid2018auxetic} and the bending and buckling of bonds at large strains. In addition, the dynamics during aging are far more complex in real systems, involving a broad range of time scales~\cite{pashine2019directed}. Nevertheless, the striking similarity between the experimental results and our numerical ones suggests that the governing  principles by which a material ages are of broader generality and independent of the precise interactions and geometry. 

Our results are also similar to the behavior seen in other more complex materials, such as rubbers~\cite{diani2009review}, solid foams~\cite{lakes_r_Science_1987} and actin networks~\cite{majumdar2018mechanical}. In the Mullins effect, a variety of rubbers undergo softening when they are strained~ \cite{bouasse1903courbes,mullins1969softening,diani2009review}. Softening occurs up to the strain to which they were deformed, similar to behavior in the k-model. Sticky colloidal gels and glasses also exhibit shear softening~\cite{delgado,koos}. The behavior we find for aging under compression in the $\ell$-model is consistent with the scenario proposed by Lakes~\cite{lakes_r_Science_1987} to explain the transformation of solid foams, that yields a negative Poisson's ratio. Furthermore, the stiffening under shear of fibrillar networks \cite{lubenskyjanmeynaturepaper,majumdar2018mechanical} is similar to our finding in the $\ell$-model; such stiffening has been explained in terms of the change in the network geometry, consistent with the underpinnings of the $\ell$-model. Thus, the effects we discuss here could apply more broadly to other disordered soft matter systems, where structure is sensitive to strain.
%%%%%%%%%%%%%%%%%%%%%%%%%%%%%%%%%%%%%%%%%%%%%

A key finding of our experiments that is also seen in our simulations is that aging imprints a memory of
strain at which the system was prepared. In the $k$-model,
the minimum of the Poisson's ratio marks the aging strain. In the
$\ell$-model the system remembers the strain that corresponds
to the initial state. The difference between these behaviors could
provide an experimental test to distinguish the dominant effects. We note that in both models, 
memory is inherently a nonlinear effect, as it is measured from the  \emph{strain dependence} of response functions. 
%the target strain must be nonlinear in order to cause aging, so the memory of the strain also occurs in the nonlinear regime.
This memory is another example of the broad range of memories that occur in out-of-equilibrium
disordered systems \cite{fiocco_PRL_2014,keim_PRL_2011,kiem_PRE_2013,kiem_Softmatter_2015,paunlsen_PRL_2014,keim_RMP_2019}. The insights gleaned from the models studied here could be relevant in those situations.

Finally, we have extended the theoretical understanding of tunability
for linear response~\cite{goodrich2015principle,hexner2018linking} to the nonlinear regime. The ability to train a response that depends on strain requires that stresses at different strains become
uncorrelated. This is quantified by measuring correlation in $B_{i}\left(\epsilon\right)$
and $G_{i}\left(\epsilon\right)$ with their corresponding value in
linear response. We find that reducing the coordination number towards the minimum threshold for mechanical stability increases
the ability to tune the system, and that it is much easier to tune the system under compression than under expansion.

We are grateful to Chukwunonso Arinze and Arvind Murugan for enlightening discussions. Work
was supported by the NSF MRSEC Program DMR-1420709 (NP) and 
DOE DE-FG02-03ER46088 (DH), DE-FG02-05ER46199 (AJL)  and the Simons Foundation for the collaboration
``Cracking the Glass Problem'' award $\#$348125 (SRN), and the Investigator award $\#$327939 (AJL). We acknowledge support from the University of Chicago Research Computing Center.

\bibliographystyle{unsrt}
\bibliography{biblo}

\begin{thebibliography}{10}

\bibitem{Struik_1977_Polymer}
LCE Struik.
\newblock Physical aging in plastics and other glassy materials.
\newblock {\em Polymer Engineering \& Science}, 17(3):165--173, 1977.

\bibitem{Hodge_1995_Science}
Ian~M Hodge.
\newblock Physical aging in polymer glasses.
\newblock {\em Science}, 267(5206):1945--1947, 1995.

\bibitem{mitchell2008aging}
James~K Mitchell.
\newblock Aging of sand--a continuing enigma?
\newblock {\em International Conference on Case Histories in Geotechnical
  Engineering}, 8, 2008.

\bibitem{Hutchinson_1995_review}
John~M Hutchinson.
\newblock Physical aging of polymers.
\newblock {\em Progress in Polymer Science}, 20(4):703--760, 1995.

\bibitem{pashine2019directed}
Nidhi Pashine, Daniel Hexner, Andrea~J Liu, and Sidney~R Nagel.
\newblock Directed aging, memory, and nature's greed.
\newblock {\em Science Advances}, 5(12):eaax4215, 2019.

\bibitem{goodrich2015principle}
Carl~P Goodrich, Andrea~J Liu, and Sidney~R Nagel.
\newblock The principle of independent bond-level response: Tuning by pruning
  to exploit disorder for global behavior.
\newblock {\em Physical review letters}, 114(22):225501, 2015.

\bibitem{lakes_r_Science_1987}
Roderic Lakes.
\newblock Foam structures with a negative poisson{\textquoteright}s ratio.
\newblock {\em Science}, 235(4792):1038--1040, 1987.

\bibitem{RLakes_Science_1987}
Roderic Lakes.
\newblock Foam structures with a negative poisson{\textquoteright}s ratio.
\newblock {\em Science}, 235(4792):1038--1040, 1987.

\bibitem{greaves2011poisson}
George~Neville Greaves, AL~Greer, Roderic~S Lakes, and Tanguy Rouxel.
\newblock Poisson's ratio and modern materials.
\newblock {\em Nature materials}, 10(11):823, 2011.

\bibitem{ren2018auxetic}
Xin Ren, Raj Das, Phuong Tran, Tuan~Duc Ngo, and Yi~Min Xie.
\newblock Auxetic metamaterials and structures: A review.
\newblock {\em Smart materials and structures}, 27(2):023001, 2018.

\bibitem{reid2019auxetic}
Daniel~R Reid, Nidhi Pashine, Alec~S Bowen, Sidney~R Nagel, and Juan~J
  de~Pablo.
\newblock Ideal isotropic auxetic networks from random networks.
\newblock {\em Soft Matter}, 15:8084--8091, 2019.

\bibitem{Lakes_2017}
Roderic~S. Lakes.
\newblock Negative-poisson's-ratio materials: Auxetic solids.
\newblock {\em Annual Review of Materials Research}, 47(1):63--81, 2017.

\bibitem{YLiu_2010}
Yanping Liu and Hong Hu.
\newblock A review on auxetic structures and polymeric materials.
\newblock {\em Scientific Research and Essays}, 5(10):1052--1063, 2010.

\bibitem{yang2004review}
Wei Yang, Zhong-Ming Li, Wei Shi, Bang-Hu Xie, and Ming-Bo Yang.
\newblock Review on auxetic materials.
\newblock {\em Journal of materials science}, 39(10):3269--3279, 2004.

\bibitem{alderson2007auxetic}
Alderson Alderson and KL~Alderson.
\newblock Auxetic materials.
\newblock {\em Proceedings of the Institution of Mechanical Engineers, Part G:
  Journal of Aerospace Engineering}, 221(4):565--575, 2007.

\bibitem{evans2000auxetic}
Kenneth~E Evans and Andrew Alderson.
\newblock Auxetic materials: functional materials and structures from lateral
  thinking!
\newblock {\em Advanced materials}, 12(9):617--628, 2000.

\bibitem{sanami2014auxetic}
Mohammad Sanami, Naveen Ravirala, Kim Alderson, and Andrew Alderson.
\newblock Auxetic materials for sports applications.
\newblock {\em Procedia Engineering}, 72:453--458, 2014.

\bibitem{saxena2016three}
Krishna~Kumar Saxena, Raj Das, and Emilio~P Calius.
\newblock Three decades of auxetics research- materials with negative poisson's
  ratio: a review.
\newblock {\em Advanced Engineering Materials}, 18(11):1847--1870, 2016.

\bibitem{liu2019realizing}
Jun Liu, Yunhuan Nie, Hua Tong, and Ning Xu.
\newblock Realizing negative poisson's ratio in spring networks with
  close-packed lattice geometries.
\newblock {\em Physical Review Materials}, 3(5):055607, 2019.

\bibitem{rens2019rigidity}
Robbie Rens and Edan Lerner.
\newblock Rigidity and auxeticity transitions in networks with strong
  bond-bending interactions.
\newblock {\em arXiv preprint arXiv:1904.07054}, 2019.

\bibitem{hexner2019periodic}
Daniel Hexner, Andrea~J Liu, and Sidney~R Nagel.
\newblock Periodic training of creeping solids.
\newblock {\em arXiv preprint arXiv:1909.03528}, 2019.

\bibitem{choi1992non}
JB~Choi and RS~Lakes.
\newblock Non-linear properties of polymer cellular materials with a negative
  poisson's ratio.
\newblock {\em Journal of Materials Science}, 27(17):4678--4684, 1992.

\bibitem{reid2018auxetic}
Daniel~R Reid, Nidhi Pashine, Justin~M Wozniak, Heinrich~M Jaeger, Andrea~J
  Liu, Sidney~R Nagel, and Juan~J de~Pablo.
\newblock Auxetic metamaterials from disordered networks.
\newblock {\em Proceedings of the National Academy of Sciences},
  115(7):E1384--E1390, 2018.

\bibitem{degarmo1997materials}
Ernest~Paul DeGarmo, J~Temple Black, Ronald~A Kohser, and Barney~E Klamecki.
\newblock {\em Materials and process in manufacturing}.
\newblock Prentice Hall Upper Saddle River, 1997.

\bibitem{WolffsLaw}
Julius Wolff.
\newblock {\em The Law of Bone Remodeling}.
\newblock Berlin Heidelberg New York: Springer, 1986 (translation of the German
  1892 edition).

\bibitem{huiskes2000effects}
Rik Huiskes, Ronald Ruimerman, G~Harry Van~Lenthe, and Jan~D Janssen.
\newblock Effects of mechanical forces on maintenance and adaptation of form in
  trabecular bone.
\newblock {\em Nature}, 405(6787):704, 2000.

\bibitem{bowden2001friction}
Frank~Philip Bowden, Frank~Philip Bowden, and David Tabor.
\newblock {\em The friction and lubrication of solids}, volume~1.
\newblock Oxford university press, 2001.

\bibitem{dieterich1994direct}
James~H Dieterich and Brian~D Kilgore.
\newblock Direct observation of frictional contacts: New insights for
  state-dependent properties.
\newblock {\em Pure and Applied Geophysics}, 143(1-3):283--302, 1994.

\bibitem{dillavou2018nonmonotonic}
Sam Dillavou and Shmuel~M Rubinstein.
\newblock Nonmonotonic aging and memory in a frictional interface.
\newblock {\em Physical review letters}, 120(22):224101, 2018.

\bibitem{maxwell1867iv}
James~Clerk Maxwell.
\newblock Iv. on the dynamical theory of gases.
\newblock {\em Philosophical transactions of the Royal Society of London},
  (157):49--88, 1867.

\bibitem{staddon2019mechanosensitive}
Michael~F Staddon, Kate~E Cavanaugh, Edwin~M Munro, Margaret~L Gardel, and
  Shiladitya Banerjee.
\newblock Mechanosensitive junction remodelling promotes robust epithelial
  morphogenesis.
\newblock {\em bioRxiv}, page 648980, 2019.

\bibitem{cavanaugh2019rhoa}
Kate~E Cavanaugh, Michael~F Staddon, Ed~Munro, Shiladitya Banerjee, and
  Margaret~L Gardel.
\newblock Rhoa mediates epithelial cell shape changes via mechanosensitive
  endocytosis.
\newblock {\em bioRxiv}, page 605485, 2019.

\bibitem{Ohern}
Corey~S. O'Hern, Leonardo~E. Silbert, Andrea~J. Liu, and Sidney~R. Nagel.
\newblock Jamming at zero temperature and zero applied stress: The epitome of
  disorder.
\newblock {\em Phys. Rev. E}, 68:011306, 2003.

\bibitem{liu2010jamming}
Andrea~J Liu and Sidney~R Nagel.
\newblock The jamming transition and the marginally jammed solid.
\newblock {\em Annu. Rev. Condens. Matter Phys.}, 1(1):347--369, 2010.

\bibitem{vanHecke_2015}
Wouter~G. Ellenbroek, Varda~F. Hagh, Avishek Kumar, M.~F. Thorpe, and Martin
  van Hecke.
\newblock Rigidity loss in disordered systems: Three scenarios.
\newblock {\em Phys. Rev. Lett.}, 114:135501, 2015.

\bibitem{bertoldi2010negative}
Katia Bertoldi, Pedro~M Reis, Stephen Willshaw, and Tom Mullin.
\newblock Negative poisson's ratio behavior induced by an elastic instability.
\newblock {\em Advanced materials}, 22(3):361--366, 2010.

\bibitem{majumdar2018mechanical}
Sayantan Majumdar, Louis~C Foucard, Alex~J Levine, and Margaret~L Gardel.
\newblock Mechanical hysteresis in actin networks.
\newblock {\em Soft matter}, 14(11):2052--2058, 2018.

\bibitem{hexner2018linking}
Daniel Hexner, Andrea~J Liu, and Sidney~R Nagel.
\newblock Linking microscopic and macroscopic response in disordered solids.
\newblock {\em Physical Review E}, 97(6):063001, 2018.

\bibitem{alderson1994auxetic}
KL~Alderson, AP~Pickles, PJ~Neale, and KE~Evans.
\newblock Auxetic polyethylene: the effect of a negative poisson's ratio on
  hardness.
\newblock {\em Acta Metallurgica et Materialia}, 42(7):2261--2266, 1994.

\bibitem{alderson2000auxetic}
Andrew Alderson, John Rasburn, Simon Ameer-Beg, Peter~G Mullarkey, Walter
  Perrie, and Kenneth~E Evans.
\newblock An auxetic filter: a tuneable filter displaying enhanced size
  selectivity or defouling properties.
\newblock {\em Industrial \& Engineering Chemistry Research}, 39(3):654--665,
  2000.

\bibitem{alderson2001auxetic}
A~Alderson, J~Rasburn, KE~Evans, and JN~Grima.
\newblock Auxetic polymeric filters display enhanced de-fouling and pressure
  compensation properties.
\newblock {\em Membrane Technology}, 2001(137):6--8, 2001.

\bibitem{diani2009review}
Julie Diani, Bruno Fayolle, and Pierre Gilormini.
\newblock A review on the mullins effect.
\newblock {\em European Polymer Journal}, 45(3):601--612, 2009.

\bibitem{bouasse1903courbes}
Henri Bouasse and Z{\'e}phyrin Carri{\`e}re.
\newblock Sur les courbes de traction du caoutchouc vulcanis{\'e}.
\newblock In {\em Annales de la Facult{\'e} des sciences de Toulouse:
  Math{\'e}matiques}, volume~5, pages 257--283, 1903.

\bibitem{mullins1969softening}
Leonard Mullins.
\newblock Softening of rubber by deformation.
\newblock {\em Rubber chemistry and technology}, 42(1):339--362, 1969.

\bibitem{delgado}
Mehdi Bouzid and Emanuela Del~Gado.
\newblock Network topology in soft gels: Hardening and softening materials.
\newblock {\em Langmuir}, 34(3):773--781, 2017.

\bibitem{koos}
Erin Koos, Wolfgang Kannowade, and Norbert Willenbacher.
\newblock Restructuring and aging in a capillary suspension.
\newblock {\em Rheologica acta}, 53(12):947--957, 2014.

\bibitem{lubenskyjanmeynaturepaper}
Cornelis Storm, Jennifer~J Pastore, Fred~C MacKintosh, Tom~C Lubensky, and
  Paul~A Janmey.
\newblock Nonlinear elasticity in biological gels.
\newblock {\em Nature}, 435(7039):191, 2005.

\bibitem{fiocco_PRL_2014}
Davide Fiocco, Giuseppe Foffi, and Srikanth Sastry.
\newblock Encoding of memory in sheared amorphous solids.
\newblock {\em Phys. Rev. Lett.}, 112:025702, Jan 2014.

\bibitem{keim_PRL_2011}
N~C Keim and Sidney~R. Nagel.
\newblock Generic transient memory formation in disordered systems with noise.
\newblock {\em Physical review letters}, 107 1:010603, 2011.

\bibitem{kiem_PRE_2013}
Nathan~C. Keim, Joseph~D. Paulsen, and Sidney~R. Nagel.
\newblock Multiple transient memories in sheared suspensions: Robustness,
  structure, and routes to plasticity.
\newblock {\em Phys. Rev. E}, 88:032306, Sep 2013.

\bibitem{kiem_Softmatter_2015}
Nathan~C. Keim and Paulo~E. Arratia.
\newblock Role of disorder in finite-amplitude shear of a 2d jammed material.
\newblock {\em Soft Matter}, 11:1539--1546, 2015.

\bibitem{paunlsen_PRL_2014}
Joseph~D. Paulsen, Nathan~C. Keim, and Sidney~R. Nagel.
\newblock Multiple transient memories in experiments on sheared non-brownian
  suspensions.
\newblock {\em Phys. Rev. Lett.}, 113:068301, Aug 2014.

\bibitem{keim_RMP_2019}
Nathan~C Keim, Joseph Paulsen, Zorana Zeravcic, Srikanth Sastry, and Sidney~R
  Nagel.
\newblock Memory formation in matter.
\newblock {\em Reviews of Modern Physics}, in press, 2019.

\end{thebibliography}

\end{document}